\documentstyle[epsf,eqsecnum,floats,preprint,aps]{revtex}

\tighten

\begin{document}

\newcommand{\Bd}{{\dot B}}
\newcommand{\Cd}{{\dot C}}
\newcommand{\fd}{{\dot f}}
\newcommand{\hd}{{\dot h}}
\newcommand{\ep}{\epsilon}
\newcommand{\vp}{\varphi}
\newcommand{\al}{\alpha}
\newcommand{\be}{\begin{equation}}
\newcommand{\ee}{\end{equation}}
\newcommand{\bea}{\begin{eqnarray}}
\newcommand{\eea}{\end{eqnarray}}
\def\gapp{\mathrel{\raise.3ex\hbox{$>$}\mkern-14mu
              \lower0.6ex\hbox{$\sim$}}}
\def\gsim{\gapp}
\def\lapp{\mathrel{\raise.3ex\hbox{$<$}\mkern-14mu
              \lower0.6ex\hbox{$\sim$}}}
\def\lsim{\lapp}
\newcommand{\PSbox}[3]{\mbox{\rule{0in}{#3}\includegraphics{#1}\hspace{#2}}}
\def\Tr{\mathop{\rm Tr}\nolimits}
\def\su#1{{\rm SU}(#1)}

\baselineskip 12pt

\title{Creation and Structure of Baby Universes in Monopole Collisions}

\author{Arvind Borde$^a$\footnote[1]{\tt borde@cosmos2.phy.tufts.edu}\footnote[2]{Permanent
address: Theoretical and Computational Studies Group,  
Natural Science Division,
Southampton College of Long Island University, 
Southampton, NY 11968, USA.}, 
Mark Trodden$^b$\footnote[3]{\tt trodden@theory1.phys.cwru.edu}
and
Tanmay Vachaspati$^b$\footnote[4]{\tt txv7@po.cwru.edu.}}

\address{~\\ $^a$Institute of Cosmology\\
Department of Physics and Astronomy\\
Tufts University\\
Medford, MA 02155, USA.}

\address{~\\$^b$Particle Astrophysics Theory Group \\
Department of Physics \\
Case Western Reserve University \\
10900 Euclid Avenue \\
Cleveland, OH 44106-7079, USA.}

\maketitle

\begin{abstract}%
Under certain circumstances, the collision of magnetic monopoles, 
topologically locked-in regions of false vacuum, leads to topological 
inflation and the creation of baby universes. 
The future evolution of initial data represented by the
two incoming monopoles may contain a timelike singularity but 
this need not be the case. We discuss the global structure
of the spacetime associated with monopole collisions and also
that of topological inflation. We suggest
that topological inflation within magnetic monopoles leads
to an eternally reproducing universe.
\end{abstract}

\setcounter{page}{0}
\thispagestyle{empty}

\vfill

\noindent CWRU-P25-98 \hfill 


\noindent LIU-SC-TCSG-1/1998001 \hfill Typeset in REV\TeX

\eject

\baselineskip 24pt plus 2pt minus 2pt

\section{Introduction}

The inflationary universe scenario provides a natural mechanism by which
an initially small region of space can expand exponentially in a short time.
The particular implementation of inflation in which we are interested is
{\it topological inflation}.  Exponential expansion occurs here
within the cores of topological defects such as monopoles, vortices, or 
domain walls~\cite{vilenkin,linde}. In this scenario, the formation of
defects in a phase transition soon after the big bang is accompanied
by inflation within the defects, provided certain parameters assume
values in suitable ranges.  We are interested in this paper in the 
dependence of topological inflation on particle physics parameters and, 
in particular, on the topological winding of
the defect, since it is
possible to change the winding by bringing together several defects.
In the case of vortices~\cite{DTV 98} 
(see also~\cite{Cho 98}), there exists a
range of parameters for which the conditions for topological inflation 
are satisfied for high winding vortices, but not for low winding ones. 
Can the conditions that are necessary for inflation 
to take place be satisfied today if we merge
small winding defects to produce a larger winding defect? If so, we could, in 
principle, create a ``baby universe'' in the laboratory. 

In general, topological inflation occurs if the width $w$ of a topological
defect is larger than the horizon size corresponding to the energy 
density $\rho_V$ in the core of the defect:

\be
w>\left(\frac{3}{8\pi G\rho_V}\right)^{1/2} \ ,
\label{infcond1}
\ee
which, for unit winding defects, is typically satisfied for symmetry
breaking scales $\eta$ larger than the Planck mass $m_p$. 
We refer to defects that do not
satisfy the inequality (\ref{infcond1}) as {\it subcritical}, 
and those that do as {\it supercritical}.

An intuitive understanding of why higher topological charges alleviate the
high symmetry breaking scales required for topological inflation
can be obtained by considering the asymptotic form of the metric 
for a static cosmic string solution aligned with the 
$z$-axis \cite{Gott}:

\be
ds^2 = dt^2 - dr^2 - dz^2 - r^2 d{\tilde\theta}^2 \ .
\label{staticmetric}
\ee
Here ($r,{\tilde\theta},z$) are cylindrical polar coordinates 
in a locally Minkowski but globally conical spatial section, 
with, for critical coupling, ${\tilde\theta}$ taking values in the range

\be
0 \leq {\tilde\theta} < 2\pi\left(1-4|n|\mu_1\right) \ ,
\ee
where $n$ is the topological winding number of the string, and 
$\mu_1 \sim \eta^2$ is the energy per unit length of a
string with unit winding (in Planck units).
This static metric is applicable as long as the deficit angle is less 
than $2\pi$ and hence static solutions cease to exist for
\be
\eta \gsim {{m_p} \over {2\sqrt{|n|}}} \ .
\label{ceases}
\ee
Thus, the critical scale at which asymptotically static 
solutions become impossible decreases with increasing winding
as $1/\sqrt{n}$. Of course, the absence of static
solutions does not guarantee that the core will inflate. A
numerical study~\cite{DTV 98} (see also~\cite{Cho 98}) shows, however, 
that topological inflation does set in at critical symmetry breaking 
scales with approximately this dependence on $n$.

These results imply that one could start with
several non-inflating $n=1$ vortices and combine them to form a large winding
vortex which would then start inflating.
In this paper we argue that a similar process holds
for magnetic monopoles.  Further, in certain models, colliding magnetic
monopoles in a regular, asymptotically flat spacetime can lead to the
creation of a region of space that satisfies all of the conditions for
inflation to occur~-- that is, it has the inflationary equation of state
in a sufficiently large spatial volume. This is our ``baby universe''.
Farhi and Guth~\cite{FG 87} have, however, pointed out that
the future
development of such a spacetime is likely to be marred by 
singularities. Their result rests on a theorem of
Penrose~\cite{P 65} that states that a final (initial) singularity
must occur in a spacetime in which there exists a trapped 
(anti-trapped) surface, as long as

\begin{enumerate}
\item $R_{\mu\nu}k^{\mu}k^{\nu} \geq 0 \ \forall$ null $k^{\mu}$
(where $R_{\mu\nu}$ is the Ricci tensor), and
\item the spacetime contains a noncompact Cauchy surface.
\end{enumerate}
A trapped (anti-trapped)
surface here is a closed 2-surface for which both the ingoing and outgoing 
sets of light rays
normal to the surface are converging (diverging) at every point on the surface.

Application of the Penrose theorem to the collision of monopoles
means that, under its assumptions, a singularity must be present to the 
past of any region
containing anti-trapped surfaces. It follows that
we are unlikely to be able to evolve initial
data to the point where we can find anti-trapped surfaces (such as
occur in de Sitter space) without first encountering a singularity.
This is the ``obstacle'' to the creation of an actual baby universe 
in the laboratory discussed by Farhi and Guth. 

Singularities can be avoided, however, in spacetimes in which the 
assumptions of Penrose's theorem are violated. In such spacetimes,  
the future evolution of the expanding baby universe is regular.
Assumption~(1) follows in Einstein's theory from the weak energy condition
(which says that the matter energy density must be positive as measured
by any observer).  The assumption is violated if we allow negative
matter energy densities, or if we look at certain alternate theories
of gravitation, in which there are extra terms in the field equation
that allow assumption~(1) to be violated even when the weak energy condition
holds.  Examples
of such theories are some higher derivative gravity models~\cite{nonsing},
and dilaton-inspired scalar-tensor models~\cite{us}. In these models the
singularity is avoided because of the existence of a limiting length, such 
as one might expect if string theory were the underlying physics.
We do not consider the violation of assumption~(1) any further in this paper.

Assumption~(2) is a very strong one and there are examples of solutions
to Einstein's equation (with reasonable matter) that violate 
it~\cite{hawkingellis}.
Without the assumption, several scenarios are possible~\cite{AB 97}.
The one that is of greatest use to us, since we are interested in creating
regions of de~Sitter space, has a closed Universe forming to the future
of the trapped surface.
This scenario will become clear in Sec.~\ref{singularity} when we draw 
the Penrose diagrams for the spacetime in which two non-inflating monopoles 
collide to form a higher winding monopole within which the conditions for 
topological inflation are met. 

In earlier work~\cite{{numerics1},{numerics2},{numerics3}} monopoles 
undergoing 
topological inflation have been studied by solving the classical
equations of motion. The initial conditions 
are chosen to mimic cosmological conditions and the numerical results 
reveal that the core of the monopole expands exponentially.
In Sec.~\ref{cosmology} we consider the
large-scale structure of spacetime for the cosmological scenario
of topological inflation within magnetic monopoles and construct
spacetime diagrams.  Several important features of these diagrams
are forced on us by general theorems on global spacetime structure. 
Some of the claims in~\cite{numerics1} are inconsistent with our diagrams.
We believe that there is an error in that work in the computation of the
behavior of null geodesics, as we explain in Sec.~\ref{cosmology}.

In Sec.~\ref{su5monopoles} we discuss the specific example of 
monopoles in an $SU(5)$ theory, and show that there is a region of parameter
space in which the unit winding monopoles do not satisfy the conditions
for topological inflation but the higher winding monopoles do. The
discussion in this section is meant to provide an example in which 
spherically symmetric monopoles of various windings can exist and be stable. 
Our results on the spacetime of inflating monopoles from the previous
sections are valid more generally.

In the concluding section we point out the
possibility that the inflating region of spacetime may detach from the
asymptotic region. If this is true, it is possible that the 
detached baby universe may split into three other universes 
which can then each split into three more, and so on, ad infinitum.
Then the monopole core may contain a very large number of baby
universes and not just one. Finally, we also
point out the possible relevance of topological inflation to models in
which there is a duality between particles and solitons.

\section{Spacetime Structure for Baby Universe Creation}
\label{singularity}

Consider the spacetime evolution as two monopoles with unit winding
($n=1$) collide and coalesce \footnote{A rich variety of gravitating
magnetic monopole solutions are known and it has been proposed that
it may be possible to understand these using catastrophe theory
\cite{tachizawaetal94}. For a given set of parameters, there can
be more than one solution to the Einstein-Yang-Mills equations,
but only one solution is stable. Here we shall always assume 
that we are working with the stable solution for any given set of 
parameters.}.
We assume that the $n=1$ monopoles are
not inflating. Hence the metric external to the $n=1$ monopoles is
Reissner-Nordstrom and can be continued smoothly within the interiors
of the monopole cores. Thus, at early times the spacetime is 
asymptotically flat and there are no singularities. When
the monopoles collide and coalesce, the result is an $n=2$ monopole,
which we assume satisfies the conditions for
topological inflation. (In Sec.~\ref{su5monopoles} we will show
that this is possible in certain models.) However, the condition
for topological inflation necessarily requires that the monopole
be a black hole~\cite{vilenkin}. To see this note that the black 
hole condition is that the width of the monopole be less than the 
associated Schwarzschild radius
\be
w\lsim 2Gm \ ,
\label{BHcond}
\ee
with $m$ the monopole mass. (We are assuming that $\rho_V$ is
large and hence the magnetic charge of the monopole is small compared 
to its mass in natural units.) Now, we may estimate 
the mass of the monopole by assuming that the core has constant energy 
density $\rho_V$, which yields $m\sim 4\pi \rho_V w^3/3$. Inserting 
this into (\ref{BHcond}) gives

\be
w \gsim
\left(\frac{3}{8\pi G\rho_V}\right)^{1/2} \ ,
\ee
which means that the condition for a monopole to undergo topological 
inflation is identical to the condition that it is a black hole as
seen from the exterior region.

Are there singularities in this spacetime?
Since the black hole conditions are satisfied, the spacetime is
likely to have trapped surfaces 
and hence it would appear from Penrose's theorem that
there must be a singularity to the future of 
the initial data (incoming $n=1$ monopoles).
Also, if the interior of the monopole is to inflate, the spacetime there
must become approximately de Sitter and, since de Sitter space has 
anti-trapped surfaces, it would again appear that there must be a
singularity on which at least one past directed null geodesic originating
in the interior terminates. Thus it would seem that there
are two singularities in the spacetime. An important point here, however,
is that the singularity that lies to the future of the asymptotic 
Reissner-Nordstrom region can be the same singularity as lies
to the past of the de Sitter region
Thus, the ``crunch'' of the 
black hole can play the role of the big bang for the inflating region.

\begin{figure}[htb]
\centerline{\epsfysize=10cm \epsfbox{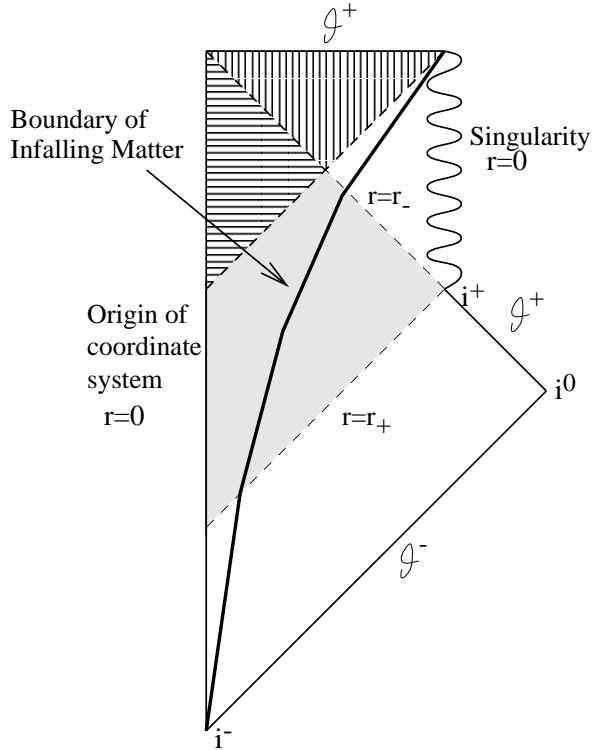}}
\vspace{0.3in}
\caption{\sf Penrose diagram for a singular 
spacetime as two subcritical $n=1$
monopoles collide and produce a supercritical $n=2$ monopole, the interior
of which then satisfies the conditions for topological inflation and
is a candidate baby universe. The supercritical monopole is necessarily
a black hole.  The surface $r=r_+$ is the event horizon. The infalling
matter lies in the region between the origin of the coordinate system
and the thick curve.}
\label{baby}
\end{figure}

In Fig.~\ref{baby} we show the spacetime structure of the inflating 
monopoles when there is an ``initial-final'' singularity of this type
\footnote{The spacetime is similar to that shown in 
Hawking and Ellis~\cite{hawkingellis} (page 361) for the collapse of 
a charged dust cloud to form a Reissner-Nordstrom black hole.}. 
Each point in the interior of this diagram represents a spacelike
2-sphere.
The horizon structure is exactly that of part of the Reissner-Nordstrom 
spacetime. In the horizontally shaded region, 
outgoing future directed null
rays escape to future null infinity (${\cal J}^+$) even though they are 
inside the monopole. The equation of state in this region is inflationary
($p=-\rho$), and we can time evolve our initial data (the well-separated
$n=1$ monopoles) to make predictions about this region. 
Note, however, that there are no anti-trapped surfaces in the 
horizontally shaded region, since those would be inconsistent with
the Penrose theorem.
The anti-trapped surfaces appear in the vertically shaded region.
Incoming and outgoing null rays directly expand here to null infinity.
To the past of this region, as suggested by Farhi and Guth, there is
a singularity. 
Predictability in this region is lost due to signals
originating at the singularity and at $i^+$. However, if boundary 
conditions at the timelike singularity and infinity can be imposed, 
predictability will be
restored. Trapped surfaces occur in the region
shaded with light gray, and the singularity to their future is the same
as the one to the past of the anti-trapped surfaces.

\begin{figure}[htb]
\centerline{\epsfysize=10cm \epsfbox{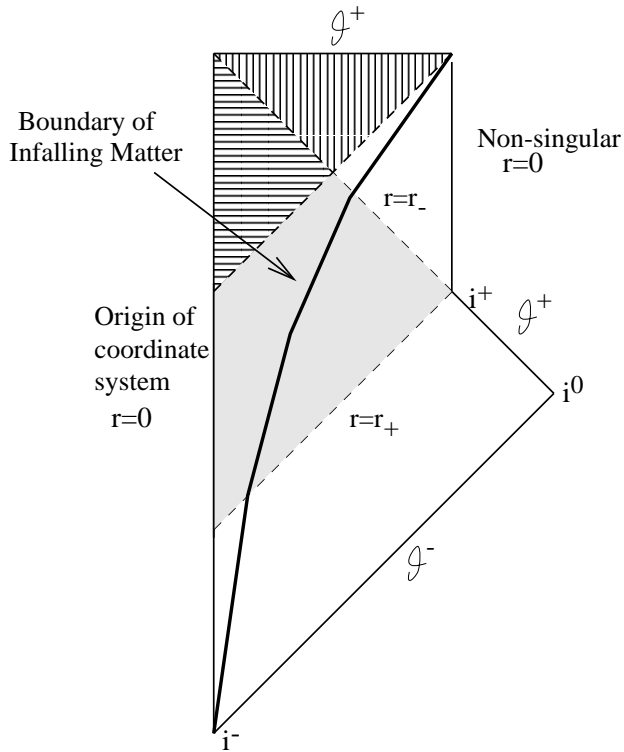}}
\vspace{0.3in}
\caption{\sf Penrose diagram for a nonsingular 
spacetime as two subcritical $n=1$ monopoles collide and produce a 
supercritical $n=2$ monopole, the interior
of which then satisfies the conditions for topological inflation and
is a baby universe. As in Fig. 1, the supercritical 
monopole is necessarily a black hole and the surface $r=r_+$ is the 
event horizon.}
\label{nosing}
\end{figure}

Is this singularity necessary?
In Einstein's theory with ``normal'' matter, the singularity
is required to occur if there is a noncompact Cauchy 
surface.  The
example in Fig.~\ref{baby} does not possess such a surface (the 
vertically shaded
region, for example, lies outside the Cauchy development of any asymptotically
flat, initial value hypersurface).  It is known that singularities can be
avoided in such scenarios~\cite{AB 97}.
We then have a spacetime structure like the one in Fig.~\ref{nosing}. 
The global structure here is similar to that of Fig.~\ref{baby}
(and the differently shaded regions have the same meanings), 
except that there are no singularities. This means that a spacelike 
slice between one $r=0$ origin of coordinates and the other $r=0$ 
line must be a 3-sphere.  In other words a closed Universe evolves
out of our initially open one. In this case  
predictability is lost due to signals originating at~$i^+$.

It has been shown under very general assumptions~\cite{AB 94,AB 97}
that these are the only two possibilities: we must either 
have singularities or we must have topology change.

\section{Cosmological Topological Inflation}
\label{cosmology}

In contrast to the monopole collision scenario
that we discussed previously, the standard picture of topological 
inflation is
that the phenomenon occurs in the extremely early universe. In this case,
it is reasonable to ask if the initial singularity that may be associated
with baby 
universe production is in fact the usual big bang singularity. If so,
the presence of anti-trapped surfaces would not require any additional
singularities. (The presence of trapped surfaces, as seen from
the asymptotic region may, of course, still lead
to a future singularity.) 

\begin{figure}[htb]
\centerline{\epsfysize=10cm \epsfbox{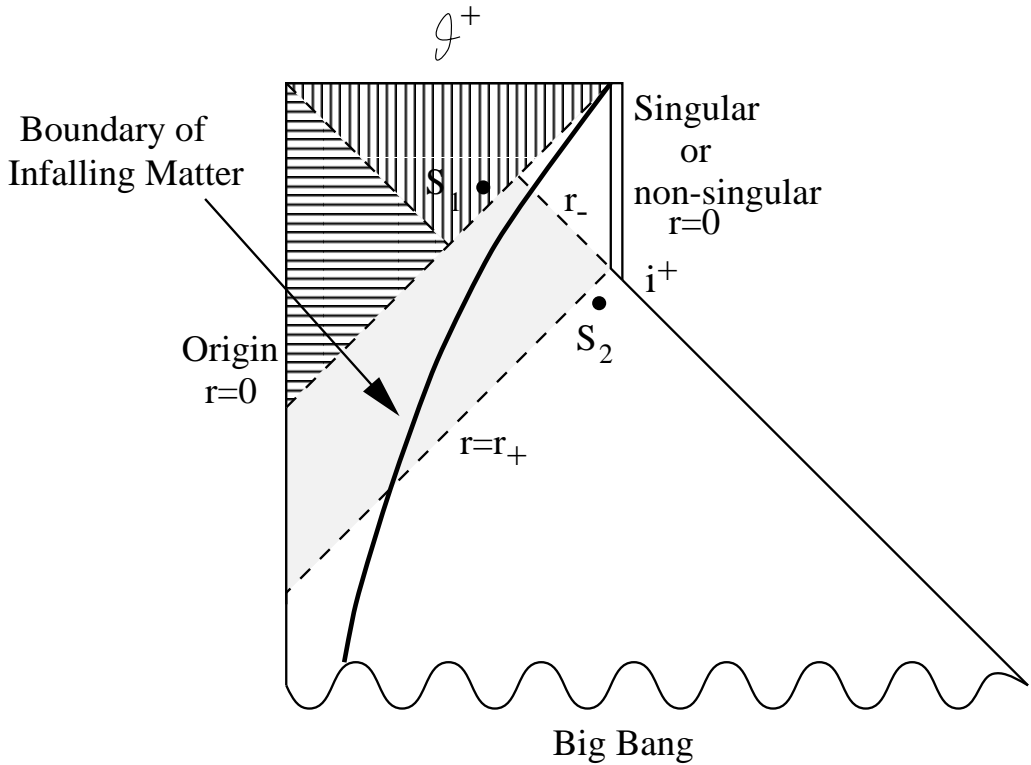}}
\vspace{0.3in}
\caption{\sf A possible Penrose diagram for the spacetime as a 
supercritical 
topological defect undergoes topological inflation in the very early 
universe. For reasons explained in the text, this is not a viable
diagram.}
\label{nonviable}
\end{figure}

A careful look at the possibilities does
not, however, seem to permit us to unify the usual big bang 
singularity with
any that might be associated with baby 
universe production. Consider, for example, an attempt to
glue a de Sitter region to the interior of a Reissner-Nordstrom
black hole between $r_-$ and $r_+$, the 
inner and outer horizons.  This will
mean that there are anti-trapped 2-surfaces in the de Sitter region
whose past does not intersect the usual Reissner-Nordstrom
singularity.  Then the only singularity that might be associated
with our baby universe is the usual big bang one.  Such
a spacetime is schematically depicted in Fig.~\ref{nonviable}.
There are trapped surfaces in the region
shaded with light gray, and de Sitter anti-trapped surfaces in the 
region with vertical shading.  (The determination of what surfaces
are trapped or anti-trapped is done via purely local calculations
of the expansion of null geodesics, and thus depends solely on the 
local metric and not on the global properties of the spacetime.) 
The double vertical line shown is the singularity that might exist 
associated with the trapped surfaces (although, topology change would
allow us to avoid this singularity, as we've mentioned earlier). 
There is also a limiting case of this scenario in which 
the de Sitter region is glued to the Reissner-Nordstrom
spacetime along $r_+$. There are no trapped surfaces here and 
no Reissner-Nordstrom singularity.

The properties of these spacetimes are, however, self-contradictory.
Consider an anti-trapped surface $S_1$ whose past does not intersect
the Reissner-Nordstrom singularity, and consider another surface
$S_2$, a ``normal'' 2-sphere just outside the horizon, to the
past of~$S_1$. We expect such surfaces $S_2$ close to the horizon to
behave as they do in the usual Reissner-Nordstrom case and to
be unaffected by cosmological considerations.
In-going light
rays from $S_2$ are thus converging, but when they reach the region
that contains $S_1$ they must be diverging without yet having focused. 
This can only occur if the weak energy condition is violated.

\begin{figure}[htb]
\centerline{\epsfysize=10cm \epsfbox{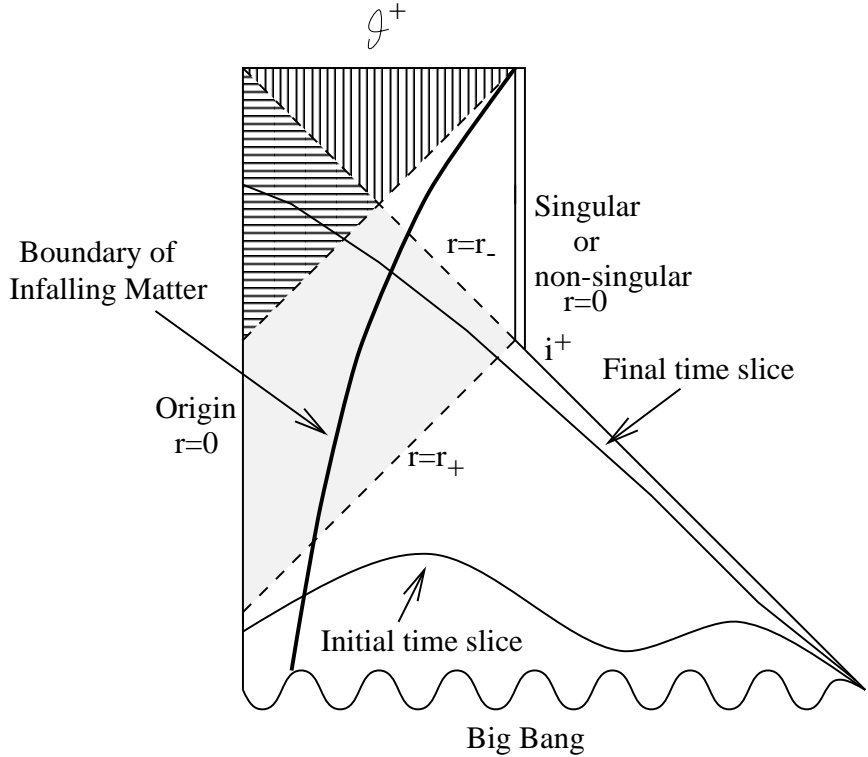}}
\vspace{0.3in}
\caption{\sf Penrose diagram for the spacetime as a supercritical 
topological defect undergoes topological inflation in the very early 
universe.}
\label{viable}
\end{figure}

The only possibility appears to be the one shown in
Fig.~\ref{viable}.
Again, there are trapped surfaces in the region
shaded with light gray, and anti-trapped surfaces in the region with
vertical shading.  The singularity shown is the one that 
might exist 
associated with both the trapped and the anti-trapped surfaces 
(although, topology change would allow you to avoid this singularity 
here as well).  There is also a separate cosmological singularity.

In Fig.~\ref{viable} we also show a spacelike hypersurface 
on which one can
specify initial data and then evolve numerically until the 
Reissner-Nordstrom singularity develops or the topology change occurs. 
Such an evolution has been studied in 
\cite{numerics1,numerics2}. 
It has been claimed in~\cite{numerics1} that the evolution can be
followed all the way until anti-trapped surfaces form. This might be 
possible if the global spacetime structure were like the one shown in
Fig.~\ref{nonviable}. But, as we have argued, such models are
not viable. In Fig.~\ref{viable} the anti-trapped region lies
outside the Cauchy development of the initial time slice, and no
final time slice can intersect it.
The horizon structure claimed in~\cite{numerics1} is also more 
complicated than ours. The horizon structure and the claim that
anti-trapped surfaces can be seen to form are both based on a 
computation of the expansion of radial null geodesics at different points
in spacetime.  There appears to be an error, however, in the computation 
of the expansion in~\cite{numerics1}, which throws into question
the entire discussion there of 
horizons~\footnote{
The metric in~\cite{numerics1} is $ds^2=-dt^2+A^2(t,r)dr^2+B^2(t,r)r^2
(d\theta^2+\sin^2\theta d\phi^2)$.  The radial null vector field
$N^{\mu}=(-1,\pm A^{-1},0,0)$ is used in that reference in order
to determine trapped and anti-trapped surfaces. But $N^{\mu}$ 
is not the tangent vector associated with
{\it affinely parametrized\/} radial null geodesics, and so 
its expansion, $\theta\equiv N^{\mu}_{\>;\mu}$, cannot be used
to find trapped and anti-trapped surfaces.}.

\section{A Specific Model: $SU(5)$ Monopoles and Topological Inflation}
\label{su5monopoles}
Consider the familiar Grand Unified symmetry breaking scheme

\be
G\equiv SU(5) \rightarrow \frac{SU(3)\times SU(2)\times U(1)}{{\cal Z}_6}
\equiv H_{SM} \ ,
\ee
realized by the vacuum expectation value (VEV) of an $SU(5)$ adjoint
Higgs field. If we write~\cite{L&V 97} the potential of the Higgs as
\be
V(\Phi)=-m_1^2[\Tr(\Phi^2)] + a[\Tr(\Phi^2)]^2 + b\Tr(\Phi^4) \ ,
\label{potential}
\ee
with $m_1$ a mass scale and $a$, $b$ dimensionless parameters, 
then the appropriate VEV is

\be
\langle\Phi\rangle = v_1{\rm diag}(2,2,2,-3,-3) \ ,
\label{VEV}
\ee
with

\be
v_1 \equiv \frac{m_1}{\sqrt{60a+14b}} \ .
\label{v1def}
\ee
It is known~\cite{GH 84,G 84} that the resulting spectrum of stable 
monopole solutions consists
of those with topological windings $n=1,2,3,4,6$. External
to any monopole solution, the symmetry group of the vacuum is $H_{SM}$. 
However, the symmetry group in the core $H^{(n)}_{core}$ depends on the 
topological charge.  We shall focus on the following cases:

\bea 
H^{(1)}_{core} & = & SU(2)\times SU(2)\times U(1)
\nonumber \\
H^{(2)}_{core} & = & SU(4)\times U(1) \ .
\eea
In the broken phase, where the gauge group is that of the standard
model, we may decompose $\Phi$ into three pieces, transforming as 
$({\bf 8},{\bf 1})$, $({\bf 1},{\bf 3})$, $({\bf 1},{\bf 1})$, representations
of $\su3 \times \su2$, with masses (following the notation of~\cite{L&V 97})

\be
\mu_8  \equiv  \sqrt{20b}\, v_1 \ , \ \ 
\mu_3  \equiv  2\mu_8 \ , \ \ 
\mu_0  \equiv  2m_1
\ee
respectively.

The potentially interesting regime for us is one in which the following 
conditions are satisfied:

\begin{enumerate}
\item The correct $\su5$ symmetry breaking occurs.
\item The $n=1$ monopoles can attract to form an $n=2$ monopole.
\end{enumerate}

These criteria are satisfied by the following parameter choices. The
symmetry breaking occurs if we choose

\be
\mu_3 = 2\mu_8 \ ,
\label{condition1}
\ee
and the $n=1$ monopoles attract for

\be
2\mu_8 > \mu_0 \ .
\label{condition2}
\ee

Now, the condition~(\ref{infcond1}) that a monopole inflates, may be 
expressed in terms of our mass parameters as

\be
\left(\frac{8\pi G \rho_V}{3}\right)^{1/2} = H > \mu_3 = 2\mu_8 \ .
\label{inflatecondition}
\ee
We are interested in the possibility that, within the parameter range we 
have specified, the $n=2$ monopoles might satisfy (\ref{inflatecondition}) 
but the $n=1$ monopoles might not. We may calculate the energy densities
$\rho_V^{(n)}$ in the cores of monopoles of winding $n$. In the regime above, 
the relevant monopoles satisfy

\bea
\rho_V^{(1)} & = & \frac{m_1^4}{4b}\left[\left(\frac{1}{x+7/30}\right)
-\left(\frac{1}{x+\alpha_1}\right)\right] \nonumber \\
\rho_V^{(2)} & = & \frac{m_1^4}{4b}\left[\left(\frac{1}{x+7/30}\right)
-\left(\frac{1}{x+\alpha_2}\right)\right] \ ,
\label{vacenergy}
\eea
where

\be
\alpha_1 = \left(\frac{8}{35}\right)^2\left(\frac{113}{16}+\frac{2}{256}\right)
\ , \ \ \ \ \alpha_2 = \frac{16}{25} \ ,
\ee
and $x\equiv a/b$. From these expressions it is easy to see that 
$\rho_V^{(1)}<\rho_V^{(2)}$ for a suitable choice of parameters. Thus,
$SU(5)$ monopoles provide a natural particle physics setting in which our
scenario could operate.

\section{Conclusions}
\label{conclusions}

We have investigated a novel mechanism for producing a baby universe
in the laboratory. In particular, we have shown that, in a certain range 
of parameters, magnetic monopoles can collide to 
produce a spacetime structure which is a Reissner-Nordstrom black hole from
the original asymptotic region, but has an interior with the inflationary
equation of state. This baby universe will begin to expand, but
the extent to which we may predict its later evolution is unclear. 
In one scenario, there exists a singularity and the future evolution of the 
baby universe is unpredictable.  In another scenario there is no 
spacetime singularity, and a closed baby universe develops. The future
evolution of the baby universe is unpredictable here as well.

\begin{figure}[htb]
\centerline{\epsfysize=8cm \epsfbox{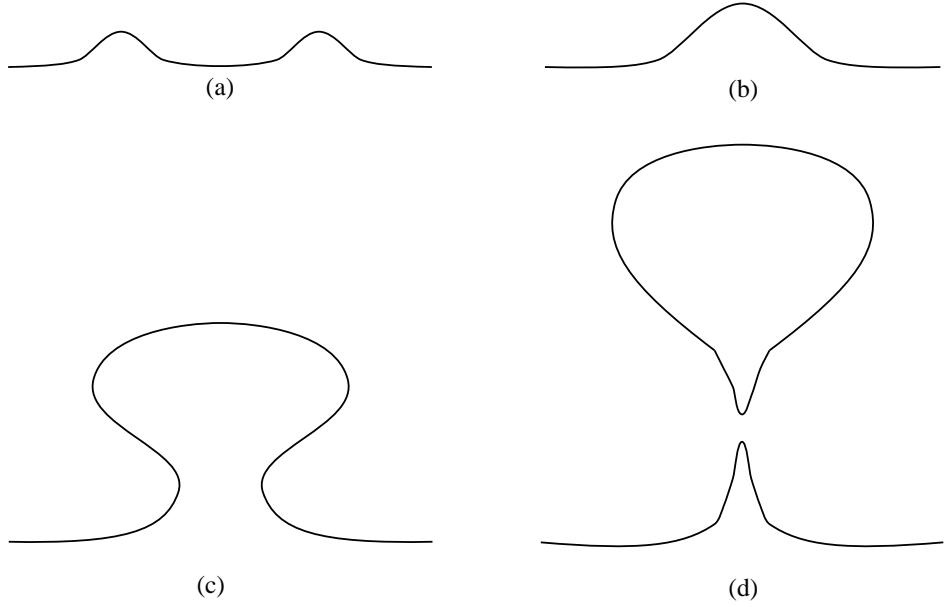}}
\vspace{0.3in}
\caption{\sf Embedding diagrams showing the incoming unit winding
monopole, their coalescence to form a higher winding, inflating
monopole, and the development of a pinch that disconnects the baby
universe from the mother universe.
}
\label{pinch}
\end{figure}

One possible development of the spacetime~\cite{BGG} 
is that the inflating region pinches off
the asymptotic spacetime, leading to a disconnected baby universe. It is
useful to picture this process using an embedding diagram. 
To construct the embedding diagram, we suppress one spatial dimension
and consider the resulting 2+1 dimensional spacetime at a fixed time $t$. 
We then construct the induced metric on a rotationally 
invariant spatial 2-surface 
$z=f(r)$, embedded in 3+1 dimensional Minkowski
space. This metric is:

$$
ds^2 = (1+{f'}^2)dr^2 + r^2 d\theta^2 
$$
where, $f'$ denotes the derivative of $f$ with respect to $r$.
For a certain choice of function $f$, and for a suitably defined
radial coordinate $R(r)$, this
metric will be identical to the truncated metric of our spacetime.
The surface $z=f(r)$ then gives the embedding diagram at different
times. The behaviour of the embedded surface with time
is shown schematically in 
Fig.~\ref{pinch}. (Parts (b) and (c) of Fig.~\ref{pinch} correspond
to the embedding diagrams of Sakai~\cite{numerics1}.) 

\begin{figure}[htb]
\centerline{\epsfysize=10cm \epsfbox{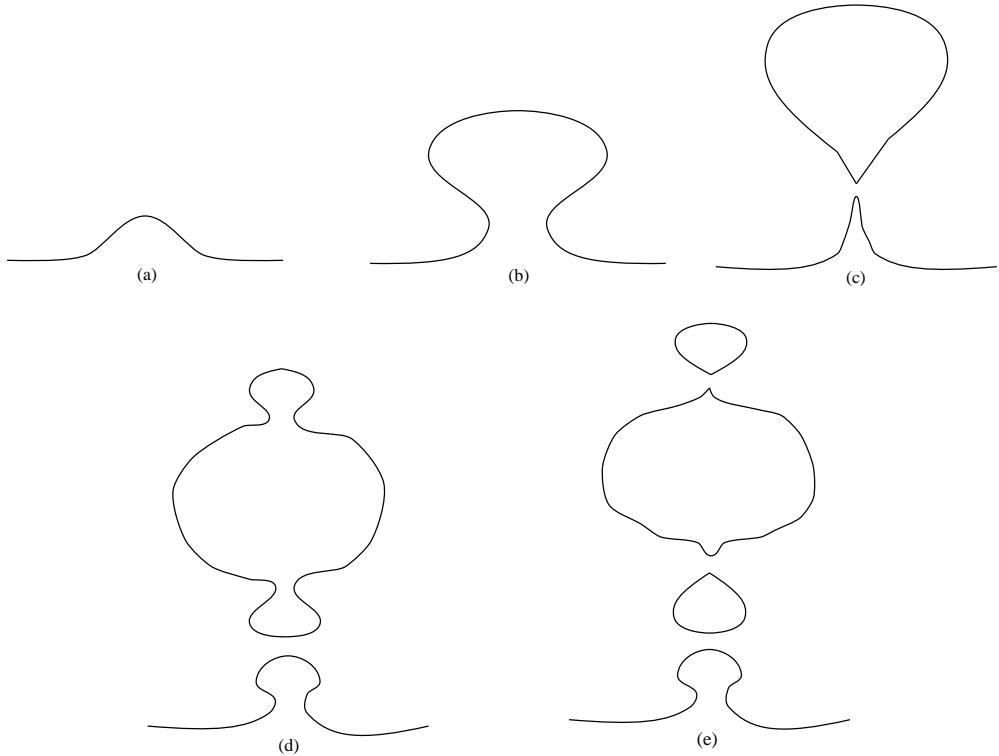}}
\vspace{0.3in}
\caption{\sf 
Embedding diagrams showing the subdivision of the baby universe.
}
\label{mitosis}
\end{figure}

The possibility that the baby universe disconnects from the original
spacetime leads to further questions. What happens to the magnetic
charge in the original spacetime? Since the asymptotic magnetic
field is left intact by the pinching off, we conclude that the
singularity must be accompanied by the pair creation of a magnetic
monopole and antimonopole. The magnetic monopole stays on at the
Reissner-Nordstrom singularity of the asymptotic spacetime while
the antimonopole is attached to the singularity at the ``south
pole'' of the detached baby universe. This is also consistent
with Gauss' law since the total magnetic charge of the closed 
(baby) universe must vanish. But now the baby universe with
a monopole at the north pole and the antimonopole at the south
pole, inflates. Eventually, the equatorial region will be inflated
far away from the polar regions. In this situation, the northern
(and southern) hemispheres are exactly like the spacetime shown
in Fig.~\ref{pinch}b. Further development must lead to inflation
of the polar cap regions,
ultimately leading to their pinching off, yielding two new
universes  in addition to the original
baby universe (see Fig.~\ref{mitosis}). Since this
process must continue forever, it leads to an eternally
reproducing universe.

To conclude, the collisions of heavy monopoles may
lead to the scenario described in this paper. However, the
creation of a baby universe cannot be observed by its creator since 
signals from within the baby universe cannot reach the asymptotic region. 

Another possibility arises if
the particles that we observe in nature are in fact the magnetic monopoles 
of another dual theory~\cite{tv,L&V 97}. In that case, it is possible 
that stellar collapse into black holes can lead to the production of 
a baby universe, since such a collapse corresponds to the coalescence of about
$10^{57}$ particles. An important issue in this scenario is that the net 
electric charge of a star is zero and so 
it corresponds to the coalescence of
monopoles and antimonopoles with zero net winding. However, if all
that is required for inflation is the occurrence of a large region
of false vacuum, even the collision of monopoles and antimonopoles
might provide the right conditions.

\acknowledgements
We thank Glenn Starkman, Cyrus Taylor and Alex Vilenkin for useful discussions.
This work was supported by the Department of Energy (D.O.E.). M.T. is also
supported by the National Science Foundation (N.S.F.).
A.B. thanks the Research Awards Committee 
of Southampton College for its financial support and the Institute
of Cosmology at Tufts University and the High Energy Theory Group at Brookhaven
National Laboratory for their hospitality.


\begin{thebibliography}{}

\bibitem{vilenkin}
A. Vilenkin, {\it Phys. Rev. Lett.} {\bf 72}, 3137 (1994).

\bibitem{linde}
A. Linde, {\it Phys. Lett.} {\bf B327}, 208 (1994).

\bibitem{DTV 98}
A. de Laix, M. Trodden and T. Vachaspati, 
{\it Phys. Rev.} {\bf D57}, 7186 (1998).

\bibitem{Cho 98}
I. Cho, ``Inflation and Nonsingular Spacetimes of Cosmic Strings'',
gr-qc/9804086. 

\bibitem{Gott}
A. Vilenkin, {Phys. Rev} {\bf D23}, 852 (1981);
J.R. Gott, {\it Ap. J.} {\bf 288}, 422 (1985);
W.A. Hiscock, {\it Phys. Rev.} {\bf D31}, 3288 (1985).

\bibitem{FG 87}
E. Farhi and A. Guth, {\it Phys. Lett.} {\bf B183}, 149 (1987).

\bibitem{P 65}
R. Penrose, {\it Phys. Rev. Lett.} {\bf 14}, 57 (1965).

\bibitem{nonsing}
V. Mukhanov and R. Brandenberger, {\it Phys. Rev. Lett.} {\bf 68}, 
1969 (1992); R. Brandenberger, V. Mukhanov and A. Sornborger, 
{\it Phys. Rev.} {\bf D48}, 1629 (1993).

\bibitem{us}
M. Trodden, V. Mukhanov, R. Brandenberger, Phys. Lett. B316 483 (1993);
R. Moessner and M. Trodden, Phys. Rev. D51, 2801 (1995).

\bibitem{hawkingellis}
S.W. Hawking and G.F.R. Ellis, ``The large-scale structure of
space-time'', Cambridge University Press (1973).

\bibitem{AB 97}
A. Borde, {\it Phys. Rev.} {\bf D55}, 7615 (1997).

\bibitem{numerics1}
N. Sakai, {\it Phys. Rev.} {\bf D54}, 1548 (1996)

\bibitem{numerics2}
N. Sakai, H. Shinkai, T. Tachizawa and  K. Maeda, {\it Phys. Rev.} {\bf D53},
655 (1996); Erratum-{\it ibid} {\bf D54}, 2981 (1996).

\bibitem{numerics3}
I. Cho, A. Vilenkin, {\it Phys. Rev.} {\bf D56}, 7621 (1997).

\bibitem{tachizawaetal94}
T. Tachizawa, K. Maeda and T. Torii, {\it Phys. Rev.} {\bf D51}, 
4054 (1995).

\bibitem{AB 94}
A. Borde, {\it Phys. Rev.} {\bf D50}, 3692, (1994).

\bibitem{GH 84}
C. Gardner and J. Harvey, {\it Phys. Rev. Lett.} {\bf 52}, 879 (1984).

\bibitem{G 84}
C. Gardner, {\it Phys. Lett.} {\bf B142}, 379 (1984).

\bibitem{L&V 97}
H. Liu and T. Vachaspati, {\it Phys. Rev.} {\bf D56}, 1300 (1997).

\bibitem{BGG}
S. Blau, E. Guendelman and A. Guth, {\it Phys. Rev.} {\bf D35}, 1747 (1987).

\bibitem{tv}
T. Vachaspati, {\it Phys. Rev. Lett.} {\bf 76}, 188 (1996).


\end{thebibliography}
\end{document}